\documentclass[aps,preprint,prb]{revtex4}
\usepackage{graphicx}
\usepackage{epsfig}
\usepackage{amsmath}
\usepackage{graphics}
\begin{document}

\author{M. F. Gelin}

\affiliation{Department of Chemistry, Technical University of Munich,  D-85747 Garching, Germany}

\author{D. S. Kosov}

\affiliation{Department of Physics and Center for Nonlinear Phenomena and Complex Systems,
Universit\'e Libre de Bruxelles, Campus Plaine, CP231, Boulevard du Triomphe,
B-1050 Bruxelles, Belgium} 

\affiliation{ Department of Chemistry and Biochemistry,
 University of Maryland, College Park, 20742, USA}

\title{Microscopic origin of the jump diffusion model}

\begin{abstract}
The present paper is aimed at studying the microscopic origin of
the jump diffusion. Starting from the $N$-body Liouville equation
and making only the assumption that molecular reorientation is overdamped,
we derive and solve the new (hereafter generalized diffusion) equation. This   
is the most general equation which governs orientational 
relaxation of an equilibrium molecular ensemble in the hindered rotation limit and in the long time limit. 
The generalized diffusion equation is an extension of the small-angle diffusion equation
beyond the impact approximation. We establish the conditions under which the generalized
diffusion equation can be identified with the jump diffusion equation, and also discuss the similarities and differences between the two approaches.    

\end{abstract}
\maketitle

\section{Introduction}

How do the molecules reorient in water? Certainly, their rotation
is significantly hindered and the orientational correlation functions
(OCFs) exhibit the long time exponential decay.
It would not be thus unreasonable to expect that the small-angle rotational diffusion would 
accurately describe reorientation of water molecules.    
On the contrary, the very first molecular dynamics simulations have revealed that water
reorientation hardly obeys the small-angle diffusion. \cite{sti71,McD82}
In fact, virtually all the simulations performed so far on liquid
water support the jump-diffusion mechanism of reorientation of water molecules. 
\cite{bag05,ber98,lit72,lad08}
Very recently, a new approach has been developed, which views rotation of water molecules
in terms of breaking and making of hydrogen bonds, and the extended
jump-diffusion model has been put forward. \cite{hyn06,hyn06a,hyn07,lud07} 

Two fundamental questions arise then: 
What is wrong with the small-angle diffusion? 
And why is the jump diffusion applicable to water reorientation? 
Indeed, the small-angle diffusion
\cite{fav60,ste63,ste84a,eis72,hub72,McCo,mor82} is well known to
be a legitimate description of molecular reorientation in the overdamped
limit, i.e. when the angular momentum is a fast variable on
the timescale of reorientation. Then the integral
relaxation time of the angular momentum correlation function yields,
through the Green-Kubo relation, the small angle rotational diffusion
coefficient. \cite{fav60,ste63,ste84a,eis72,hub72,McCo,mor82} For liquid water,
the integral angular momentum relaxation time ($\sim$ a few femtoseconds)  
is much shorter than the  orientational relaxation time ($\sim$ several picoseconds), 
so that the small-angle diffusion should perform excellently. It does not, however.

The jump diffusion \cite{val73,cuk72,cuk74,sil99,lep01a}  
is formulated through the master equation
in the space of orientations. The ensuing OCFs are specified by the two phenomenological parameters,  
the averaged jump angle and the jump rate. The parameters do not have any molecular origin or specificity.  
They can hardly be traced back or even related to moments of inertia or rotational friction. The very notions of angular momentum and rotational relaxation are alien to the jump diffusion. 
It is thus surprising that such 
a model, which is  normally used for a phenomenological description of molecular reorientations is solids and glasses, is applicable to liquid water.

The present paper is aimed at studying the microscopic origin of
the jump diffusion and explaining the failure of the small-angle diffusion.
Starting from the $N$-body Liouville equation
and making only the assumption that molecular reorientation is overdamped
(a precise meaning of this requirement is concretized in Sec. 4),
we derive and solve the new (hereafter generalized diffusion) equation.
 This is the most general equation which governs orientational 
relaxation of an equilibrium molecular ensemble in the hindered rotation limit and in the long time limit.
Similarly to the small-angle and jump diffusion,  the generalized
diffusion equation predicts exponentially decaying OCFs.
The generalized diffusion equation is an extension of the small-angle diffusion equation
beyond the impact approximation.
 It can be rewritten as the small-angle diffusion
equation, in which the diffusion coefficients depend explicitly on
the rank $j$ of the OCF. 
 We establish the conditions under which the generalized
diffusion equation can be identified with the jump diffusion equation, 
and also discuss the similarities and differences between the two approaches.

\section{generalized master equation}

We start with a formally exact Zwanzig-type master equation, which
can be derived from the $N$-particle rotation-translational
Liouville equation by applying the projection operator technique \cite{fre75,eva78,gel98}

\begin{equation}
\partial_{t}\rho(\mathbf{J},\mathbf{\Omega},t)=-i\Lambda(\mathbf{J},\mathbf{\Omega})\rho(\mathbf{J},\mathbf{\Omega},t)-\int_{0}^{t}dt'\Re(\mathbf{J},\partial_{\mathbf{J}},\hat{\mathbf{L}}(\mathbf{\Omega}),t-t')\rho(\mathbf{J},\mathbf{\Omega},t').\label{kin1}\end{equation}
 Here $\rho(\mathbf{J},\mathbf{\Omega},t)$ is the single particle
probability distribution, $\mathbf{J}$ is the angular momentum
in the molecular frame, $\hat{\mathbf{L}}(\mathbf{\Omega})$ is the
angular momentum operator in the molecular frame, $\mathbf{\Omega}$
denotes collectively the set of three Euler angles $\alpha,\beta,\gamma$
that specify orientation of the molecular frame with respect to the
laboratory one. The free-rotor Liouville operator consists of the
two contributions, \begin{equation}
\Lambda(\mathbf{J},\mathbf{\Omega})=\Lambda_{\mathbf{\Omega}}+\Lambda_{\mathbf{J}},\label{str}\end{equation}
 which describe, respectively, the angular momentum driven reorientation
and the angular momentum change during free rotation: \begin{equation}
\Lambda_{\mathbf{\Omega}}=\sum_{a=x,y,z}I_{a}^{-1}J_{a}\hat{L}_{a}(\mathbf{\Omega}),\,\,\,\Lambda_{\mathbf{J}}=-i\sum_{a,b,c=x,y,z}\varepsilon_{abc}I_{b}^{-1}J_{a}J_{b}\partial_{J_{c}},\label{str1}\end{equation}
 $I_{\alpha}$ are the main moments of inertia, $\varepsilon_{abc}$
is the Levi-Civita symbol. $\Lambda_{\mathbf{J}}\equiv0$ for linear
and spherical rotors.

The relaxation operator $\Re$ can explicitly be written as the generalized
Fokker-Planck operator \cite{fre75,eva78,gel98} \begin{equation}
\Re(\mathbf{J},\partial_{\mathbf{J}},\hat{\mathbf{L}}(\mathbf{\Omega}),t)=\sum_{a=x,y,z}\partial_{J_{a}}\Xi_{a}(\mathbf{J},\partial_{\mathbf{J}},\hat{\mathbf{L}}(\mathbf{\Omega}),t)(\partial_{J_{a}}+J_{a}I_{a}^{-1}),\label{FPgen}\end{equation}
$\Xi_{a}(\mathbf{J},\partial_{\mathbf{J}},\hat{\mathbf{L}}(\mathbf{\Omega}),t)$
being a friction operator (for a standard rotational Fokker-Planck
equation, $\Xi_{a}(\mathbf{J},\partial_{\mathbf{J}},\hat{\mathbf{L}}(\mathbf{\Omega}),t)=\delta(t)\xi_{a}$,
where $\xi_{a}$ is the constant friction). As is clear from Eq. (\ref{FPgen}),
the relaxation operator obeys the normalization

\begin{equation}
\int d\mathbf{J}\Re(\mathbf{J},\partial_{\mathbf{J}},\hat{\mathbf{L}}(\mathbf{\Omega}),t)=0\label{norm}\end{equation}
 (note that integration over $\mathbf{\Omega}$ is not necessary)
and the detailed balance \begin{equation}
\Re(\mathbf{J},\partial_{\mathbf{J}},\hat{\mathbf{L}}(\mathbf{\Omega}),t)\rho_{eq}(\mathbf{J})=0.\label{DetBal}\end{equation}
 $\rho_{ne}(\mathbf{J})$ is the equilibrium Boltzmann distribution

\begin{equation}
\rho_{eq}(\mathbf{J})=(2\pi k_{B}T)^{-3/2}(I_{x}I_{y}I_{z})^{-1/2}\exp\{-\sum_{a=x,y,z}J_{a}^{2}/(2k_{B}TI_{a})\}.\label{Boltz}\end{equation}
$k_{B}$ is the Boltzmann constant, $T$ is the temperature.

Due to the isotropy of space, the relaxation operator $\Re(\mathbf{J},\partial_{\mathbf{J}},\hat{\mathbf{L}}(\mathbf{\Omega}),t)$
cannot depend on the Euler angles $\mathbf{\Omega}$ explicitly.\cite{foott7}
However, it can
explicitly contain the angular momentum operators. \cite{fre75,eva78,gel98,rid69}
The bulk majority of the theories of molecular rotation (except the
jump-diffusion models \cite{val73,cuk72,cuk74,sil99,lep01a} and some
more general approaches \cite{fre75,eva78,gel98,rid69}), adopt the
impact approximation, which assumes that $\Re$ is independent of
$\hat{\mathbf{L}}(\mathbf{\Omega})$. \cite{BurTe,kos06a} 
As will be clear from the following discussion, retaining the $\hat{\mathbf{L}}(\mathbf{\Omega})$-dependence
in the relaxation operator is essential for getting beyond the small-angle diffusion.

\section{Orientational correlation functions}

If we expand the probability density on the Wigner D-matrices of the
rank $j$, \cite{var89}\begin{equation}
\rho(\mathbf{J},\mathbf{\Omega},t)=\sum_{j=0}^{\infty}\frac{2j+1}{8\pi^{2}}\sum_{k,l=-j}^{j}\rho_{kl}^{j}(\mathbf{J},t)D_{kl}^{*j}(\mathbf{\Omega}),\label{Green1}\end{equation}
 we arrive at the equation \begin{equation}
\partial_{t}\rho^{j}(\mathbf{J},t)=-i(\Lambda_{\mathbf{\Omega}}^{j}+\Lambda_{\mathbf{J}})\rho^{j}(\mathbf{J},t)-\int_{0}^{t}dt'\Re^{j}(\mathbf{J},\partial_{\mathbf{J}},\mathbf{L}^{j},t-t')\rho^{j}(\mathbf{J},t').\label{kin2}\end{equation}
 Here $\Lambda_{\mathbf{\Omega}}^{j}$ and $\Re^{j}$ are determined
by Eqs. (\ref{str}) and (\ref{str1}), in which the 
angular momentum
operators $\hat{L}_{a}$ are replaced by their matrix elements $L_{a}^{j}$
over the D-matrices: \begin{equation}
(L_{x}^{j})_{kl}\pm i(L_{y}^{j})_{kl}=\delta_{k,l\mp1}\{(j\pm l)(j\mp l+1)\}^{1/2},\,\,(L_{z}^{j})_{kl}=l\delta_{kl};\,\,-j\leq k,l\leq j.\label{Jxyz}\end{equation}
 In Eq. (\ref{kin2}) and below, we use the compact notation, regarding
operators $\Lambda_{\mathbf{\Omega}}^{j}$, $\Re^{j}$ and the probability
density $\rho^{j}(t)$ as $(2j+1)\times(2j+1)$ matrices, so that
the product $\Lambda_{\mathbf{\Omega}}^{j}\rho^{j}$ and similar quantities
are to be understood as the matrix products.

If the impact approximation is used, then  
the relaxation operator is $j$-independent, \begin{equation}
\Re^{j}=\Re^{0}.\label{Imp}\end{equation}
Since the relaxation operator $\Re$ is, in general,  $\hat{\mathbf{L}}(\mathbf{\Omega})$-dependent, 
 rotational and orientational relaxation is described
by a collection of rank-dependent operators $\Re^{j}$. Despite the $j$-dependence of $\Re^{j}$ is normally ignored, 
its very presence is not
entirely unexpected, since OCFs of different ranks are affected by
the effective $j$-dependent cage potentials. \cite{pol97,pol04}
Thus a popular librational oscillator model of Lynden-Bell and Steele
requires $j$-dependent values of the librational frequencies and
mean torques for reproducing simulated OCFs even for simple liquids.
\cite{ste84} Dielectric friction, which governs orientational relaxation
in polar systems, is also known to be rank-dependent. \cite{bag94}
Furthermore, the quantum master equations are subdivided into $j$-dependent
sub-operators even within the impact approximation. \cite{BurTe}

OCF of the rank $j$ can be calculated through $\rho^{j}(t)$ as follows:
\begin{equation}
\left\langle P_{j}(\mathbf{u}_{1}(0)\mathbf{u}_{2}(t)\right\rangle \equiv\sum_{k,l=-j}^{j}D_{0k}^{j}(0,-\alpha_{2},-\beta_{2})\rho_{kl}^{j}(t)D_{l0}^{j}(\alpha_{1},\beta_{1},0).\label{OCF1}\end{equation}
 Here $P_{j}$ is the Legendre polynomial, $\alpha_{i},\beta_{i}$
are the spherical angles of the unit vectors $\mathbf{u}_{i}$ ($i=1,2$)
in the molecular frame, \begin{equation}
\rho_{kl}^{j}(t)\equiv\int d\mathbf{J}\rho_{kl}^{j}(\mathbf{J},t),\,\,\,\rho^{j}(\mathbf{J},t=0)=\rho_{eq}(\mathbf{J}).\label{Ocft0}\end{equation}
 If we wish to follow reorientation of the unit vector pointing along
the molecular $z$-axis, then $\alpha_{i}=\beta_{i}=0$ and 
\begin{equation}
\left\langle P_{j}(\mathbf{u}_{z}(0)\mathbf{u}_{z}(t)\right\rangle =\rho_{00}^{j}(t).\label{zz}
\end{equation}

\section{Overdamped limit and generalized diffusion}
Taking the Laplace transform of Eq. (\ref{kin2}), we get the equivalent
equation\begin{equation}
-\rho_{eq}(\mathbf{J})+s\widetilde{\rho}^{j}(\mathbf{J},s)=-i(\Lambda_{\mathbf{\Omega}}^{j}+\Lambda_{\mathbf{J}})\widetilde{\rho}^{j}(\mathbf{J},s)-\widetilde{\Re}^{j}\widetilde{\rho}^{j}(\mathbf{J},s)\label{kin3}\end{equation}
 (hereafter, all the Laplace-transformed operators are denoted by
tilde, viz. $\tilde{f}(s)=\int_{0}^{\infty}dt\exp\{-st\} f(t)$ for
$\forall$ $f(t)$). Now we are in a position to introduce the projection
operators\begin{equation}
P=\rho_{eq}(\mathbf{J})\int d\mathbf{J}...,\,\,\,\, Q=1-P.\label{Proj}\end{equation}
 Evidently, $P\widetilde{\Re}^{j}=\widetilde{\Re}^{j}P=0$ due to
the normalization (\ref{norm}) and detailed balance (\ref{DetBal}),
respectively. Applying $P$ and $Q$ to Eq. (\ref{kin3}) and making
use of the identities $P\Lambda_{\mathbf{J}}=\Lambda_{\mathbf{J}}P=0$,
we obtain the following exact equation for $\widetilde{\rho}^{j}(s)=\rho_{eq}^{-1}(\mathbf{J})P\widetilde{\rho}^{j}(\mathbf{J},s)$:\begin{equation}
\widetilde{\rho}^{j}(s)=\{ s+\widetilde{M}^{j}(s)\}^{-1}\label{OCF2}\end{equation}
 with \begin{equation}
\widetilde{M}^{j}(s)=\int d\mathbf{J}\Lambda_{\mathbf{\Omega}}^{j}\{ s+iQ\Lambda_{\mathbf{\Omega}}^{j}+i\Lambda_{\mathbf{J}}+\widetilde{\Re}^{j}(s)\}^{-1}\Lambda_{\mathbf{\Omega}}^{j}\rho_{eq}(\mathbf{J})\}.\label{M}\end{equation}
 Since rotation is hindered, it is natural to assume that the streaming
operators can be neglected as compared to the relaxation operator,
$\widetilde{\Re}^{j}(s)$: \begin{equation}
||Q\Lambda_{\mathbf{\Omega}}^{j}||,\,||\Lambda_{\mathbf{J}}||\ll||\widetilde{\Re}^{j}(s)||\label{SR}\end{equation}
 (here $||...||$ is a suitably defined operator norm). Then, making
use of the explicit form of the streaming operator $\Lambda_{\mathbf{\Omega}}^{j}$
(\ref{str}), we can write\begin{equation}
\widetilde{M}^{j}(s)=k_{B}T\sum_{a=x,y,z}\frac{(L_{a}^{j})^{2}}{I_{a}}\widetilde{C}_{J,a}^{j}(s),\label{M1}\end{equation}
\begin{equation}
\widetilde{C}_{J,a}^{j}(s)=\frac{\left\langle J_{a}\{ s+\widetilde{\Re}^{j}(s)\}^{-1}J_{a}\right\rangle }{\left\langle J_{a}^{2}\right\rangle }.\label{CJj}\end{equation}
 The quantity $C_{J,a}^{(j)}(t)$ can be termed as the generalized
angular momentum correlation function. It yields the standard angular momentum
correlation function for $j=0$. Within the impact approximation (\ref{Imp}), $\widetilde{\Re}^{j}$
is $j$-independent, so that $C_{J,a}^{(j)}(t)$ for different $j$
are all the same and coincide with $C_{J,a}^{(0)}(t)$. In such a
case, Eq. (\ref{OCF2}) describes
the diffusion equation with memory. \cite{ste84a,BurTe,key72,kiv88}

If we are interested in the long-time behavior of OCFs, we can neglect the 
non-Markovian effects, substitute $\widetilde{\Re}^{j}(s)$ by $\widetilde{\Re}^{j}(0)$,
and invert Eq. (\ref{OCF2}) into the time domain. If the memory operator
is defined via Eqs. (\ref{M1}) and (\ref{CJj}), we then arrive at
the generalized diffusion formula\begin{equation}
\rho^{j}(t)=\exp\{-\widetilde{M}^{j}(0)t\}\equiv\exp\{-\sum_{a=1}^{3}(L_{a}^{j})^{2}\mathfrak{\mathcal{D}}_{a}^{j}t\}.\label{dif}\end{equation}
 Here $\mathfrak{\mathcal{D}}_{a}^{j}$ are the $j$-dependent generalized
diffusion constants, which are uniquely determined by the generalized
angular momentum integral relaxation times $\tau_{J,a}^{j}$: \begin{equation}
\mathfrak{\mathcal{D}}_{a}^{j}\equiv\tau_{J,a}^{j}\frac{k_{B}T}{I_{a}},\,\,\,\tau_{J,a}^{j}\equiv\widetilde{C}_{J,a}^{j}(0)\equiv\frac{\left\langle J_{a}\{\widetilde{\Re}^{j}(0)\}^{-1}J_{a}\right\rangle }{\left\langle J_{a}^{2}\right\rangle }\equiv\int_{0}^{\infty}dtC_{J,a}^{(j)}(t).\label{dif1}\end{equation}
 Eq. (\ref{dif1}) can be coined as the generalized Green-Kubo relation. 

If $\widetilde{\Re}^{j}(0)$ is $j$-independent (that is, the impact approximation (\ref{Imp}) holds), then 
the small-angle diffusion is recovered from Eq. (\ref{dif}):
\begin{equation}
\rho^{j}(t)=\exp\{-\sum_{a=1}^{3}(L_{a}^{j})^{2}\mathfrak{\mathcal{D}}_{a}t\},\label{Small}\end{equation}
\begin{equation}
\mathfrak{\mathcal{D}}_{a}\equiv\tau_{J,a}^{(0)}\frac{k_{B}T}{I_{a}}.\label{Smalla}\end{equation}
A close similarity between Eqs. (\ref{dif}) and (\ref{Small}) is
evident. A major difference is the following: the generalized angular
momentum relaxation times $\tau_{J,a}^{j}$ are not necessary equaled
to the angular momentum relaxation times $\tau_{J,a}^{(0)}$ any longer.
Same is true about the diffusion coefficients $\mathfrak{\mathcal{D}}_{a}^{j}$
and $\mathfrak{\mathcal{D}}_{a}$. Thus, the small-angle rotational
diffusion is not applicable in the overdamped limit if the $\hat{\mathbf{L}}(\mathbf{\Omega})$-dependence
of the relaxation operator
$\Re(\mathbf{J},\partial_{\mathbf{J}},\hat{\mathbf{L}}(\mathbf{\Omega}),t)$ cannot be ignored.

The generalized (\ref{dif}) and  small-angle (\ref{Small}) diffusion equations  
allow us to analytically calculate OCFs (\ref{zz}) of the first ($j=1$) and second ($j=2$) rank 
for general asymmetric top molecules (see Appendix A).

\section{Jump diffusion model}

The jump diffusion model \cite{val73,cuk72,cuk74,sil99,lep01a} can
be retrieved from the general master equation (\ref{kin1}), provided
we chose the angular momentum independent relaxation operator \begin{equation}
\Re(\mathbf{J},\partial_{\mathbf{J}},\hat{\mathbf{L}}(\mathbf{\Omega}),t)=\delta(t)\nu\Re_{\mathbf{\Omega}}(\hat{\mathbf{L}}(\mathbf{\Omega}))\label{j1}\end{equation}
($\nu$ being the jump rate) or if we consider any operator $\Re_{\mathbf{\Omega}}(\hat{\mathbf{L}}(\mathbf{\Omega}),t)$
with a finite memory in the long time limit. The jump diffusion operator
(\ref{j1}) obeys the detailed balance\begin{equation}
\Re_{\mathbf{\Omega}}(\hat{\mathbf{L}}(\mathbf{\Omega}))1=0\label{Ja}\end{equation}
(an isotropic distribution must be an eigenvector of $\Re_{\mathbf{\Omega}}$) 
and normalisation 

\begin{equation}
\int d\mathbf{\Omega}\Re_{\mathbf{\Omega}}(\hat{\mathbf{L}}(\mathbf{\Omega}))=0.\label{Jb}
\end{equation}
Note that the detailed balance (\ref{Ja}) and normalization (\ref{Jb})
conditions for operator $\Re_{\mathbf{\Omega}}$ differ from their
counterparts (\ref{norm}) and (\ref{DetBal}) for the true relaxation
operator $\Re(\mathbf{J},\partial_{\mathbf{J}},\hat{\mathbf{L}}(\mathbf{\Omega}),t)$.
Therefore, the ensuing behaviours of the OCFs are different,
too. Indeed, if we apply the hindered rotation limit (\ref{SR}) to
Eq. (\ref{kin1}) with the relaxation operator (\ref{j1}), we 
can simply neglect the streaming operator 
(\ref{str})  as comapred to the relaxation operator. Then, if we 
expand  Eq. (\ref{Green1}) over the Wigner
matrices, we get the jump diffusion OCF\begin{equation}
\rho^{j}(t)=\exp\{-\nu\Re_{\Omega}^{j}t\}.\label{jump}\end{equation}
 Here $\Re_{\Omega}^{j}$ is the matrix element of the operator $\Re_{\Omega}$
over the Wigner matrices. It is frequently assumed that $\Re_{\Omega}^{j}$
can be parametrized through the averaged rotational matrix, \begin{equation}
\Re_{\Omega}^{j}=1-\int d\mathbf{g}\rho(\mathbf{g})\exp\{-i\sum_{a=x,y,z}g_{a}L_{a}^{j}\}.\label{Ju1}\end{equation}
The modulus and direction of $\mathbf{g}$ determine the angle and
axis of rotation, $\rho(\mathbf{g})$ is the corresponding probability
density, and $L_{a}^{j}$ are given by Eq. (\ref{Jxyz}). The jump diffusion
model reduces to the small-angle diffusion in the limit of small angular
jumps ($\rho(\mathbf{g})$ is nonzero for $|g|\ll1$). In case of
isotropic jumps ($\rho(\mathbf{g})=\rho(|\mathbf{g}|)$) we get, approximately,
\cite{val73,cuk72,cuk74,sil99,lep01a,hyn06} \begin{equation}
\Re_{\Omega}^{j}=1-\frac{1}{2j+1}\frac{\sin((j+1/2)g)}{\sin(g/2)},\label{Ju2}\end{equation}
$g$ being an averaged jump angle. 

\section{Generalized diffusion vs. jump diffusion}

We now establish the similarities and differences between  the generalized diffusion and the jump diffusion.
Both of the models predict exponentially decaying OCFs. The decay
rates are governed by the relaxation matrices $\widetilde{M}^{j}(0)$
(\ref{dif1}) and $\nu\Re_{\Omega}^{j}$ (\ref{Ju1}) which, in general,
differ from the small-angle diffusion tensor (\ref{Smalla}). One
might thus prematurely conclude that both the generalized and the
jump diffusion models result in the same predictions, and the use
of any of the two is just a matter of taste. That is not the case,
however. There are several important differences between the two approaches. 

(i). The generalized diffusion tensor $\widetilde{M}^{j}(0)$ is inversely
proportional to the dissipation strength. It is given, in fact, by
the inverse of the relaxation operator $\widetilde{\Re}^{j}(0)$ (\ref{dif1}).
The jump diffusion tensor $\nu\Re_{\Omega}^{j}$ is proportional to the
jump rate $\nu$. This means that increase of dissipation (e.g.,
increase of molecular density) slows down the OCF decay
in the generalized diffusion model but speeds up the OCF decay in
the jump diffusion model.

(ii). The rank-dependence of $\widetilde{M}^{j}(0)$ is determined
by the relaxation operator $\Re(\mathbf{J},\partial_{\mathbf{J}},\hat{\mathbf{L}}(\mathbf{\Omega}),t)$.
It can be rather complicated, and it is not guaranteed that the functional
form of $\widetilde{M}^{j}(0)$ can successfully be approximated by
the jump-diffusion matrix $\Re_{\Omega}^{j}$ (\ref{Ju1}) or (\ref{Ju2}). 

(iii). $\widetilde{M}^{j}(0)$ is explicitly determined by the angular
momentum relaxation through the $j$-dependent generalized angular
momentum relaxation times $\tau_{J,a}^{j}$ (\ref{dif1}). Furthermore,
$\widetilde{M}^{j}(0)$ depends explicitly on the molecular moments of inertia 
and on temperature. No such
information is contained in the jump relaxation matrix $\Re_{\Omega}^{j}$
without additional ad hoc assumptions.

To better illustrate the similarities and differences between the generalized
diffusion and the jump diffusion, we consider below two representative examples.

\subsection{Liquid of spherical molecules}

Starting from the $N$-(spherical) particle Liouville equation, Evans has 
obtained the $\hat{\mathbf{L}}(\mathbf{\Omega})$-dependent corrections
to the relaxation operator by expanding $\Re(\mathbf{J},\partial_{\mathbf{J}},\hat{\mathbf{L}}(\mathbf{\Omega}),t)$
in powers of the (projected) streaming operator $\Lambda_{\mathbf{\Omega}}$
(\ref{str}) up to the second order. \cite{eva78}  Having utilized several other
approximations, he has derived the explicit expression for the relaxation
operator, which in our notation reads \begin{equation}
\widetilde{M}^{j}(0)=\frac{j(j+1)\tau_{J}}{I/(k_{B}T)+(j(j+1)-1/2)\tau_{J}^{2}+j(j+1)\tau_{T}^{2}},\label{Ev}\end{equation}
$\tau_{T}$ being the torque relaxation time. This expression can
equivalently be rewritten as $\widetilde{M}^{j}(0)=j(j+1)D^{j}$,
where $D^{j}=\tau_{J}^{(j)}k_{B}T/I$ is the $j$-dependent diffusion
coefficient and the generalized angular momentum relaxation time is
defined as \begin{equation}
\tau_{J}^{(j)}=\frac{\tau_{J}}{1+(k_{B}T/I)[(j(j+1)-1/2)\tau_{J}^{2}+j(j+1)\tau_{T}^{2}]}.\label{EvJ}\end{equation}

If $\sqrt{I/(k_{B}T)}\gg\tau_{J},\,\tau_{T}$, then we recover the
small-angle diffusion (\ref{Small}) with the rank-independent
diffusion coefficient (\ref{Smalla}). Now we can try to recast Eq.
(\ref{Ev}) in terms of the jump diffusion model relaxation operator
$\nu\Re_{\Omega}^{j}$. The result is quite obvious in the opposite
limit of $\sqrt{I/(k_{B}T)}\ll\tau_{J}\,\mathrm{or}\,\tau_{T}$.\cite{foott2} In
such a case, the generalized diffusion relaxation operator becomes,
approximately, $j$ and temperature independent, and we recover the so-called large-angle
jump limit of Eqs. (\ref{jump})-(\ref{Ju2}): \begin{equation}
\nu\approx\frac{\tau_{J}}{\tau_{J}^{2}+\tau_{T}^{2}},\,\,\, g_{j}\approx\frac{2\pi}{2j+1},\,\,\,\Re_{\Omega}^{j}\approx1,\label{JuL}\end{equation}
so that the effective mechanism of molecular reorientation is via
large-amplitude jumps. However, to make a connection with the jump
diffusion model, we have to introduce a $j$-dependent averaged angle
$g_{j}$ (\ref{JuL}). In all the intermediate situations ($\sqrt{I/(k_{B}T)}\sim\tau_{J}\sim\tau_{T}$)
it is quite problematic to recast Eq. (\ref{Ev}) into the jump diffusion
relaxation operator form (\ref{jump}) and to make a clear partitioning
between the jump rate $\nu$ and the jump matrix $\Re_{\Omega}^{j}$.
Anyway, even if we do so, then $\nu$ and $\Re_{\Omega}^{j}$
become temperature,  $\tau_{J}$, and $\tau_{T}$-dependent.
In a sense, such a dynamic information is alien to the jump diffusion model
and must be incorporated into it ad hoc. Summarizing, the differences in predictions of 
the generalized diffusion model and the jump diffusion model are expected to be pronounced for liquids, 
in which the angular momentum relaxation time is  comparable with the corresponding torque
relaxation time.

\subsection{Liquid water}

Svishchev and Kusalik  \cite{kus94} have performed room-temperature molecular dynamics simulations 
of SPC/E liquid water and calculated the small-angle diffusion coefficients 
$\mathfrak{\mathcal{D}}_{a}$ and orientational relaxation times 
$\tau^{j}_{a}$ of the first and second rank. For convenience,
their data are collected in Table I. 

 If we wish to interpret the data in terms of the generalized 
diffusion model, we can recalculate  $\mathfrak{\mathcal{D}}_{a}$ into 
the angular momentum relaxation times $\tau^{0}_{J,a}$ via Eq. (\ref {Smalla}), as well as  
to recalculate $\tau^{j}_{a}$ 
into the generalized angular momentum relaxation 
times $\tau^{j}_{J,a}$ via Eqs. (\ref {ORD}) and  (\ref{dif1}). The results are presented in Table II.
Clearly, the water 
reorientation does not obey the small-angle diffusion: If this were the case, 
all $\tau^{j}_{J,a}$ ($j=0,1,2$) would be the same. In reality, 
the generalized angular momentum relaxation times   $\tau^{1}_{J,a}$ and $\tau^{2}_{J,a}$
are three-four times smaller than the corresponding 
angular momentum relaxation times   $\tau^{0}_{J,a}$. Thus,  
the small-angle diffusion predicts that water molecules reorient 
much faster than they do in reality. Interestingly,   
$\tau^{1}_{J,a} \approx \tau^{2}_{J,a}$ for $a=x, z$.

According to the data of  Table I, 
$\tau^{1}_{y}/\tau^{2}_{y}=2.9$. Given this result alone, one could prematurely 
assume that reorientation of water molecules around their axes of the intermediate moment of inertia 
obeys the small angle diffusion. In reality, such a reorientation process has nothing to do with 
the small angle diffusion, since the angular momentum relaxation time $\tau^{0}_{J,y}$ exceeds the 
generalized angular momenta relaxation times 
$\tau^{1}_{J,y}$ and $\tau^{0}_{J,y}$ by the factor of $2.6$ and $3.8$, correspondingly 
(see Table II). This is a nice illustration 
of how misleading can be uncritical application of the small-angle diffusion 
beyond its domain of validity.

We can also try to interpret the results of Svishchev and Kusalik  \cite{kus94}
in terms of the jump diffusion model. As is clear from Table I and as
is confirmed by other computer simulations  \cite{ber98,lad08,ski01} and
 NMR experiments, \cite{ski01} 
reorientation of water molecules is significantly anisotropic.
Therefore, it is impossible to describe all six orientational relaxation times 
$\tau^{j}_{a}$ ($a=x, y, z$; $j=1, 2$) by a single set of the jump angle $g$ (Eq. (\ref{Ju2})) and rate $\nu$. 
The extensions of the jump-diffusion model developed in Refs.  
\cite{sil99,lep01a} make it possible to describe anisotropic jumps. 
For our purposes, it is more convenient 
to introduce the axis-dependent jump angles $g_{a}$, $a=x, y, z$. \cite{lad08}
Given essential nonsphericity of  water molecules and anisotropy of their hydrogen bonding network, 
this assumption is not unrealistic. For each $a=x, y, z$, we should find  $g_{a}$ and $\nu$
which, according to Eqs. (\ref{jump}) and (\ref{OR}), fit  $\tau^{j}_{a}$ from  Table I 
via the expression $\tau^{j}_{a}=\{\nu\Re_{\Omega}^{j}\}^{-1}$. The results of this procedure are 
presented in Table III. The jump rate $\nu$, by its definition, should be 
the same for all  $\tau^{j}_{a}$. As is seen from Table III, it is impossible to meet this requirement
because we get the axis-dependent rates  $\nu_{a}$. According to the generalized jump-diffusion  
model, we should additionally take into  account  the slow reorientation of the 
O...O vector of the pair of H-bonded water molecules. \cite{hyn06,hyn06a,hyn07} 
This appears to make the differences in $\nu_{a}$ smaller, but the problem remains.\cite{lad08} 
Therefore, interpretation and explanation of the anisotropy of water reorientation 
in terms of the jump diffusion encounters significant difficulties.

The above results do not allow us to ascertain that the generalized diffusion model is superior 
over the jump diffusion model in the interpretation of reorientation of water molecules.
 Indeed, the generalized 
diffusion model does not allow us, within itself, 
to calculate the generalized angular momentum relaxation times $\tau^{j}_{J,a}$.  
On the other hand, the use
of the (generalized) jump diffusion model for the explanation of  rotational dynamics of water molecules 
has a  microscopic justification since both $\nu$
and $\Re_{\Omega}^{j}$ can {\it independently} be extracted from molecular
dynamics simulations, at least for the OH bond reorientation. \cite{hyn06,hyn06a} Furthermore, 
numerical values of the rotational diffusion 
coefficients and orientational relaxation times presented in Table I should not be taken
as ultimate benchmark data, since different water force 
fields predict different values of the quantities.\cite{ber98}   
However, the  results of the present section show clearly that the generalized diffusion model
allows us to get some insight into the process of reorientation of water molecules. 

\section{Conclusion}

We have derived the generalized diffusion equation (\ref{dif}), which
is uniquely obtained from the many-particle Liouville equation in
the overdamped rotation limit and in the long time limit. A precise meaning of the {}``overdamped
limit'' is given by inequalities (\ref{SR}). Eq. (\ref{dif}) predicts
the exponentially decaying OCFs and reduces to the small-angle diffusion
equation (\ref{Small}) provided that the relaxation operator $\Re(\mathbf{J},\partial_{\mathbf{J}},\hat{\mathbf{L}}(\mathbf{\Omega}),t)$
is independent of the angular momentum operators $\hat{\mathbf{L}}(\mathbf{\Omega})$.
In general, $\Re$ is $\hat{\mathbf{L}}(\mathbf{\Omega})$-dependent,
and the small-angle diffusion does not hold. 
Eq. (\ref{dif}) is the most general equation which governs orientational 
relaxation of an equilibrium molecular ensemble in the hindered rotation limit and in the long time limit.

Any deviation from
the small-angle diffusion behavior indicates the breakdown of the
impact approximation (\ref{Imp}). In such a case,
the  $j$-dependence of  $\widetilde{M}^{j}(0)$ is determined by the interparticle
interaction potential and is not established by the present analysis. However, we can get 
specific predictions if we assume that the deviation from the impact approximation is small. 
For the conceptual clarity, let us consider spherical molecules. Then we can expand the generalized 
angular momentum relaxation times $\tau_{J}^{j}$ around $j=0$ and write  
\begin{equation}
\widetilde{M}^{j}(0)=j(j+1) \frac{k_{B}T}{I} \tau_{J}^{0}[1+\lambda j(j+1)]+O(\lambda^{2}),\label{Msp}\end{equation}
 $\lambda$ being a certain dimensionless, small, and $j$-independent parameter. The insertion of Eq. (\ref{Msp})  into (\ref{dif}) gives us the generalized diffusion formula, and the ratio of the first- and second-rank orientational relaxation times is predicted to be 
 $3+12\lambda+O(\lambda^{2})$.

The generalized diffusion equation (\ref{dif})  can be mapped into the jump diffusion equation  
 (\ref{jump}) if we identify $\widetilde{M}^{j}(0)$ (Eq. (\ref{M1})) 
 with the jump diffusion matrix $\nu\Re_{\Omega}^{j}$ (Eqs. (\ref{Ju1}) and  (\ref{Ju2})). 
That can unequivocally
be done in particular cases
of small jumps and large-amplitude 
jumps. For example, Eq. (\ref{Msp}) can be reproduced within the jump diffusion model 
if we put $\tau_{J}^{0}k_{B}T/I=\nu(g^{2}/6)(1+g^{2}/60)$ and $\lambda=-g^{2}/20$.
In all intermediate situations, the use of the jump diffusion
model requires additional justification, as is discussed in detail Sec. VI . Pragmatically speaking, if the actual OCFs of different ranks are well reproduced via Eqs. (\ref{jump})-(\ref{Ju2}) with $j$-independent jump rates  $\nu$ and angles $g$  (as seems to be roughly the case for the reorientation of OH bond of a water molecule \cite{hyn06,hyn06a}) then the jump diffusion model is a legitimate description for the long time exponential decays of OCFs. If the fitted $g$ and (notably)  $\nu$ have pronounced $j$-dependences, the use of the jump diffusion is dubious.      

\appendix 
\section{OCFs of asymmetric top molecules within small-angle and generalized diffusion models}

Within the small-angle diffusion model (\ref{Small}), 
OCFs (\ref{zz}) of the first ($j=1$) and second ($j=2$) rank 
are evaluated analytically for asymmetric top molecules \cite{fav60,eis72}
\begin{equation}
\left\langle P_{1}(\mathbf{u}_{z}(0)\mathbf{u}_{z}(t)\right\rangle =\rho_{00}^{1}(t)=\exp\{-(\mathfrak{\mathcal{D}}_{x}+\mathfrak{\mathcal{D}}_{y})t\}.\label{difj1}\end{equation}
\begin{equation}
\left\langle P_{2}(\mathbf{u}_{z}(0)\mathbf{u}_{z}(t)\right\rangle =\rho_{00}^{2}(t)=g_{+}\exp\{-(6\mathfrak{\mathcal{D}}+2\Delta)t\}+g_{-}\exp\{-(6\mathfrak{\mathcal{D}}-2\Delta)t\}.\label{difj2}\end{equation}
Here the following parameters are introduced: \begin{equation}
\Delta=\sqrt{\mathfrak{\mathcal{D}}_{x}^{2}+\mathfrak{\mathcal{D}}_{y}^{2}+\mathfrak{\mathcal{D}}_{z}^{2}-\mathfrak{\mathcal{D}}_{x}\mathfrak{\mathcal{D}}_{y}-\mathfrak{\mathcal{D}}_{y}\mathfrak{\mathcal{D}}_{z}-\mathfrak{\mathcal{D}}_{z}\mathfrak{\mathcal{D}}_{x}},\label{par1}\end{equation}
\begin{equation}
\mathfrak{\mathcal{D}}=(\mathfrak{\mathcal{D}}_{x}+\mathfrak{\mathcal{D}}_{y}+\mathfrak{\mathcal{D}}_{z})/3,\,\,\, g_{\pm}=\frac{2\Delta\pm3(\mathfrak{\mathcal{D}}-\mathfrak{\mathcal{D}}_{z})}{4\Delta}.\label{par2}\end{equation}
Eqs. (\ref{difj1}) and (\ref{difj2})
can (approximately) be recast into the spherically-symmetric form
\begin{equation}
\left\langle P_{j}(\mathbf{u}_{z}(0)\mathbf{u}_{z}(t)\right\rangle
 =\exp\{-j(j+1)\overline{\mathfrak{\mathcal{D}}}_{z}t\},\label{difSa}\end{equation}
where $\overline{\mathfrak{\mathcal{D}}}_{z}=(\mathfrak{\mathcal{D}}_{x}+\mathfrak{\mathcal{D}}_{y})/2$.
Eqs. (\ref{difj1}) and (\ref{difSa}) are identical for $j=1$. For $j=2$, 
numerics show that the approximate Eq. (\ref{difSa}) delivers
OCFs which are virtually indistinguishable
from their exact counterparts (\ref{difj2}). 

OCFs $\left\langle P_{j}(\mathbf{u}_{x}(0)\mathbf{u}_{x}(t)\right\rangle $
and $\left\langle P_{j}(\mathbf{u}_{y}(0)\mathbf{u}_{y}(t)\right\rangle $
are obtainable from Eqs. (\ref{difj1}), (\ref{difj2}), and (\ref{difSa}) by the cyclic permutation of 
indexes $x,y,z$. 
Evidently, the generalized diffusion OCFs are also calculated 
via Eqs. (\ref{difj1})-(\ref{difSa}), if we replace the diffusion 
coefficients $ {\mathcal{D}}_{a}$ by their rank-dependent counterparts  $ {\mathcal{D}}_{a}^{j}$ (\ref{dif1}).

The orientational relaxation times are defined as 
\begin{equation}
\tau^{j}_{a}=\int^{\infty}_{0}dt\left\langle P_{j}(\mathbf{u}_{a}(0)\mathbf{u}_{a}(t)\right\rangle.
 \label{OR}\end{equation}
According to Eq. (\ref{difSa}),
\begin{equation}
\tau^{j}_{a}=\{j(j+1)\overline{\mathfrak{\mathcal{D}}}^{j}_{a}\}^{-1}.
 \label{OR1}\end{equation}
If all three $\tau^{j}_{x},\, \tau^{j}_{y}, \, \tau^{j}_{z}$ are known, we can use  
Eq. (\ref{OR1}) to calculate the generalized diffusion coefficients through the formula 
  \begin{equation}
{\mathcal{D}}_{x}^{j}=\frac{1}{j(j+1)} 
\left \{ \frac{1}{\mathfrak \tau^{j}_{y}} + \frac{1}{\mathfrak \tau^{j}_{z}} -
 \frac{1}{\mathfrak \tau^{j}_{x}} \right\}. 
 \label{ORD}\end{equation}
${\mathcal{D}}_{y}^{j}$ and ${\mathcal{D}}_{z}^{j}$ are obtained from Eq. (\ref{ORD})
by the cyclic permutation of $x,y,z$.

\begin{acknowledgments}
This work was partially supported by the American Chemical Society
Petroleum Research Fund Grant 44481-G6 and by NSF-MRSEC Contract 
No. DMR0520471 at the University of Maryland.
\end{acknowledgments}

\newpage

\begin{table}
\caption{Small-angle difussion coefficients $\mathfrak{\mathcal{D}}_{a}$ (in ps$^{-1}$) and 
integral orientational relaxation times $\tau_{a}^{(j)}$ (in ps) simulated in Ref. \cite{kus94}
for SPC/E water at a room temperature. 
The projections $a=x, y, z$ correspond to axis of the small, intermediate, and large moments of inertia
of the water molecule.} 

\textcompwordmark{}

\begin{tabular}{ccccc}
\hline
$a$ & $x$ &  $y$  & $z$   \\
\hline
$\mathfrak{\mathcal{D}}_{a}$ & $0.46$ & $0.45$ & $0.22$    \\
$\tau_{a}^{(1)}$ & $4.46$ & $4.54$ & $2.90$    \\
$\tau_{a}^{(2)}$ & $2.00$ & $1.57$ & $1.17$  \\
\hline
\end{tabular}
\end{table}

\begin{table}
\caption{The generalized diffusion model. 
Angular momentum relaxation times $\tau_{J,a}^{(0)}$ (in fs) and 
generalized angular momentum relaxation times $\tau_{J,a}^{(1)}$, $\tau_{J,a}^{(2)}$ (in fs) 
calculated from the data of Table I via Eqs. (\ref {Smalla}) and (\ref {ORD}).
} 

\textcompwordmark{}

\begin{tabular}{ccccc}
\hline
$a$ & $x$ &  $y$  & $z$   \\
\hline
$\tau_{J,a}^{(0)}$ & $1.13$ & $2.07$ & $1.55$    \\
$\tau_{J,a}^{(1)}$ & $0.42$ & $0.80$ & $0.35$    \\
$\tau_{J,a}^{(2)}$ & $0.40$ & $0.55$ & $0.33$  \\
\hline
\end{tabular}
\end{table}

\begin{table}
\caption{The jump diffusion model.  
Jump times $\nu^{-1}$ (in ps) and jump angles $g_{a}$ (in degrees) 
calculated from the data of Table I as is explained in the text.
} 

\textcompwordmark{}

\begin{tabular}{ccccc}
\hline
$a$ & $x$ &  $y$  & $z$   \\
\hline
$\nu^{-1}$ & $1.87$ & $0.24$ & $0.85$    \\
$g_{a}$ & $68$ & $23$ & $56$    \\
\hline
\end{tabular}
\end{table}


\begin{thebibliography}{10}
\bibitem{foott7}This assumption may break down if the system is exposed to anisotropic
external fields. \cite{cuk74}
\bibitem{foott2} It is not guaranteed that Eq. (\ref{Ev}) is valid in the limit.
\end{thebibliography}
\end{document}